\documentstyle[elsart12,osa,epsf,eqsecnum,cite]{revtex}
\begin{document}


\voffset1.5cm


\title{Medium-Induced Gluon Radiation off Massive Quarks Fills the Dead Cone}
\author{N\'estor Armesto, Carlos A. Salgado and Urs Achim Wiedemann}

\address{Theory Division, CERN, CH-1211 Geneva 23, Switzerland}
\date{\today}
\maketitle

\begin{abstract}
We calculate the transverse momentum dependence of the 
medium-induced gluon energy distribution radiated off massive 
quarks in spatially extended QCD matter. In the absence of a medium, 
the distribution shows a characteristic mass-dependent depletion of 
the gluon radiation for angles $\theta < m/E$, the so-called dead
cone effect. Medium-modifications of this spectrum are calculated 
as a function of quark mass $m$, initial quark energy $E$, in-medium 
pathlength and density. Generically, medium-induced gluon radiation
is found to fill the dead cone, but it is reduced at large gluon 
energies compared to the radiation off light quarks. We quantify
the resulting mass-dependence for momentum-averaged quantities 
(gluon energy distribution and average parton energy loss), 
compare it to simple approximation schemes and discuss its 
observable consequences for nucleus-nucleus collisions at RHIC 
and LHC. In particular,
our analysis does not favor the complete disappearance of 
energy loss effects from leading open charm spectra at RHIC.
\end{abstract}

\section{Introduction}
\label{sec1}

Hadronic jets accompanying heavy quarks $c,b$ differ
from light quark and gluon initiated jets. These
differences can be attributed to the suppression
of gluon bremsstrahlung from massive charges 
at small angles $\theta < m/E$, the dead cone 
effect~\cite{Dokshitzer:fc}.
Observable consequences of this mass-dependence of the
partonic fragmentation pattern include the softening of the
light hadron spectra accompanying heavy quark 
jets~\cite{Ackerstaff:1998hz}, and the significant hardening 
of the leading charmed~\cite{Buskulic:1993iu} or 
beauty~\cite{Abbiendi:2002vt} hadron.
Mass-dependent dead cone conditions are implemented
in the modified leading logarithmic approximation which accounts
for jet multiplicity distributions~\cite{Schumm:1992xt}. 
They are also implemented in modern Monte Carlo 
simulations~\cite{Gieseke:2003rz} which provide 
a probabilistic implementation of the perturbative part of the 
parton fragmentation process in the vacuum.

How is this parton fragmentation modified if the produced
high-energy parent quark propagates through dense QCD matter, 
as is the case in ultrarelativistic nucleus-nucleus collisions 
at RHIC and at the LHC? As a first step towards addressing this
question~\cite{Gyulassy:1993hr}, several 
groups~\cite{Baier:1996sk,Zakharov:1997uu,Wiedemann:2000za,Gyulassy:2000er,Wang:2001if} calculated to leading order in energy and for an arbitrary
number of medium-induced momentum exchanges the modifications to the  
$q \to q\, g$ splitting process for energetic light quarks. These 
calculations indicate that medium effects can
result in a significant additional energy degradation of the
leading hadron which grows approximately linear with the
density of the medium and approximately quadratic with the
in-medium pathlength. Recent 
measurements~\cite{Adcox:2001jp,Adler:2003au,Adler:2002xw,Adams:2003kv,Back:2003qr,Arsene:2003yk}
of high-$p_\perp$ hadroproduction and its
centrality dependence in Au+Au collisions at $\sqrt{s_{\rm NN}} = 200$ GeV 
provide the first evidence~\cite{Wang:2003aw} for this 
medium-induced parton energy loss.
For light quarks and gluons, the formalism was also extended
to the angular dependence of the medium-modified gluon radiation
~\cite{Wiedemann:2000tf,Baier:1999ds,Baier:2001qw,Salgado:2003gb}.
This allows to discuss medium modifications of jet shape and jet 
multiplicity observables~\cite{Salgado:2003rv}.

For massive quarks, the formalism of medium-induced gluon
radiation is not developed to the same extent. 
Dokshitzer and Kharzeev~\cite{Dokshitzer:2001zm} 
suggested that the dead cone effect also reduces the medium-induced 
gluon radiation, thus resulting in a smaller suppression of leading 
charmed and beauty hadrons. They estimated 
this effect by multiplying the medium-induced gluon spectrum for 
massless quarks with a transverse momentum averaged suppression 
factor given in Eq. (\ref{4.16}) below. However, the combination
of vacuum-induced and medium-induced gluon energy distributions
is known to differ significantly from a simple 
superposition~\cite{Wiedemann:2000tf} due to the non-trivial
interplay of interference effects and elastic scattering.
Hence, it is conceivable that the medium-induced part of the gluon 
radiation differs significantly from this averaged dead cone 
approximation. Going beyond this approximation may also be
needed to distinguish mass-dependent final state effects from
non-linear modifications of the initial gluon distribution
for which open charm production may be a sensitive 
probe~\cite{Eskola:2003fk,Gelis:2003vh,Kharzeev:2003sk,Accardi:2003be}.

This motivates to parallel for the massive case the 
calculations of medium-induced gluon radiation which exist for
the massless case. To this end, two groups
~\cite{Djordjevic:2003qk,Djordjevic:2003be,Djordjevic:2003zk,Zhang:2003wk}
presented detailed calculations of parton energy loss
for massive quarks. However, these calculations are limited to the 
average energy loss of massive quarks only. Here, we go beyond these
studies i) by providing the first analysis of the double differential 
medium-induced gluon distribution as 
a function of transverse momentum and gluon energy and ii) 
by comparing this gluon distribution to the massless limit for 
a wide parameter range in quark mass, in-medium pathlength and 
medium density. In Section~\ref{sec2} we set up the path-integral
formalism for parton energy loss. In Sections~\ref{sec3} and ~\ref{sec4},
we summarize the result of our numerical calculations and we provide 
simple analytical arguments for how the mass-dependence shows up in 
the medium-induced gluon radiation. The main results and their
relevance for heavy quark production in nucleus-nucleus collisions
at RHIC and LHC are discussed in the Conclusions.

\section{Medium-induced Gluon Radiation off Massive Quarks: Formalism}
\label{sec2}

The gluon distribution radiated
from a massive quark traversing spatially extended QCD matter
can be written as~\cite{Wiedemann:2000za,Wiedemann:2000tf}
\begin{eqnarray}
  \omega\frac{dI}{d\omega\, d{\bf k}_\perp}
  &=& {\alpha_s\,  C_F\over (2\pi)^2\, \omega^2}\,
    2{\rm Re} \int_{0}^{\infty}\hspace{-0.3cm} dy_l
  \int_{y_l}^{\infty} \hspace{-0.3cm} d\bar{y}_l\,
  e^{i \bar{q} (y_l - \bar{y}_l)}
   \int d{\bf u}\, e^{-i{\bf k}_\perp \cdot {\bf u}}   \,
  e^{ -\frac{1}{2} \int_{\bar{y}_l}^{\infty} d\xi\, n(\xi)\,
    \sigma({\bf u}) }\,
  \nonumber \\
  && \times {\partial \over \partial {\bf y}}\cdot
  {\partial \over \partial {\bf u}}\,
  \int_{{\bf y}=0={\bf r}(y_l)}^{{\bf u}={\bf r}(\bar{y}_l)}
  \hspace{-0.5cm} {\cal D}{\bf r}
   \exp\left[ i \int_{y_l}^{\bar{y}_l} \hspace{-0.2cm} d\xi
        \frac{\omega}{2} \left(\dot{\bf r}^2
          - \frac{n(\xi) \sigma\left({\bf r}\right)}{i\, \omega} \right)
                      \right]\, .
    \label{2.1}
\end{eqnarray}
Here, $\omega$ and ${\bf k}_\perp$ denote the energy and transverse 
momentum of the emitted gluon, respectively. The Casimir factor
$C_F = \frac{4}{3}$ determines the coupling strength of this gluon 
to the massive quark. For numerical results, we fix the coupling
constant to $\alpha_s = 1/3$ unless stated otherwise. 
Eq.~(\ref{2.1}) resums the effects 
of arbitrary many medium-induced scatterings to leading order in $1/E$. 
Properties of the medium enter (\ref{2.1}) via the product 
of the time-dependent density $n(\xi)$ of scattering centers times 
the strength of a single elastic scattering $\sigma({\bf r})$. 
This dipole cross section $\sigma({\bf r})$ is given 
in terms of the elastic high-energy 
cross section $\vert a({\bf q})\vert^2$ of a single scatterer
in the colour octet representation,
\begin{eqnarray}
 \sigma({\bf r}) = 2 \int \frac{d{\bf q}}{(2\pi)^2}\,
                    \vert a({\bf q})\vert^2\, 
                    \left(1 - e^{-i{\bf q}\cdot {\bf r}}\right)\, .
 \label{2.2}
\end{eqnarray}
A detailed discussion of 
(\ref{2.1}) including the physical interpretation of the {\it internal}
integration variables ($y_l$, $\bar{y}_l$, ${\bf y}$, ${\bf u}$,
$\xi$) can be found in Ref.~\cite{Wiedemann:2000za}. 

The only mass-dependence of the gluon distribution (\ref{2.1}) comes
from the phase factor $\exp \left[i \bar{q} (y_l - \bar{y}_l)\right]$,
where $\bar{q}$ is defined as the difference between the total
three momentum of the initial quark ($p_1$), and the final
quark ($p_2$) and gluon ($k$),
\begin{equation}
  \bar{q} = p_1 - p_2 - k \simeq \frac{x^2\, m^2}{2 \omega}\, ,
  \qquad x = \frac{\omega}{E}\, .
  \label{2.3}
\end{equation}
For the abelian case, the same phase factor is known to give 
the mass-dependence of medium-induced photon radiation to 
leading order in $x \ll 1$, see 
Ref.~\cite{Zakharov:1996fv,Wiedemann:1999fq}. Paralleling the derivation of 
Ref.~\cite{Wiedemann:2000za} for massive quarks, we have
checked explicitly that this phase is the only mass dependence
of the gluon distribution (\ref{2.1}).

In the absence of medium effects, the gluon energy distribution
(\ref{2.1}) reduces to the unperturbed radiation $I^{\rm vac}$ 
associated to the production of a massive quark in the vacuum. We
write the full radiation spectrum as the sum of this vacuum component
and its medium-modification $I^{\rm med}$, 
\begin{eqnarray}
  \omega\frac{dI}{d\omega\, d{\bf k}_\perp}
  = \omega\frac{dI^{\rm vac}}{d\omega\, d{\bf k}_\perp}
    + \omega\frac{dI^{\rm med}}{d\omega\, d{\bf k}_\perp}\, .
  \label{2.4}
\end{eqnarray}
By construction, both the full gluon distribution 
$\omega\frac{dI}{d\omega\, d{\bf k}_\perp}$, as well as the
vacuum component $\omega\frac{dI^{\rm vac}}{d\omega\, d{\bf k}_\perp}$
have to be positive for all values of $k_\perp$ and $\omega$.
In contrast, the medium-induced modification 
$\omega\frac{dI^{\rm med}}{d\omega\, d{\bf k}_\perp}$
can be negative in some part of phase space: negative values 
correspond to a medium-induced depletion of the vacuum
component. 

In what follows, we calculate the medium-induced gluon distribution
(\ref{2.1}) in two different approximations. 
We limit the discussion to the case of a static medium 
with in-medium pathlength $L$ for which
\begin{equation}
   n(\xi) = n_0\, \Theta(L-\xi)\, .
   \label{2.5} 
\end{equation}
From the analysis of Ref.~\cite{Salgado:2002cd}, we expect 
that the case of an expanding medium can be reformulated 
in terms of a static medium (\ref{2.5}) with suitably adjusted 
density $n_0$.

\section{Opacity expansion}
\label{sec3}
The opacity expansion of the gluon distribution (\ref{2.1}) 
amounts to an expansion of the integrand 
of (\ref{2.1}) in powers of $\left[ n(\xi)\, \sigma({\bf r})\right]^N$. 
Technical details of this expansion 
can be found in Appendix~\ref{appa} and in 
Refs.~\cite{Wiedemann:2000za,Gyulassy:2000er}. 

\subsection{$N=0$ vacuum term: the dead cone effect}
\label{sec3a}
In the absence of medium effects, $n(\xi) = 0$,  the gluon distribution 
(\ref{2.1}) reduces to the zeroth order in opacity,
\begin{equation}
   \omega\frac{dI^{\rm vac}}{d\omega\, d{\bf k}_\perp}
  \equiv \omega\frac{dI(N=0)}{d\omega\, d{\bf k}_\perp}
  = \frac{\alpha_s\, C_F}{\pi^2}
     H({\bf k}_\perp)\, ,
  \label{3.1}
\end{equation}
where $H({\bf k}_\perp)$ denotes the radiation term associated
to the {\it hard} parton production,
\begin{equation}
     H({\bf k}_\perp) =
     \frac{{\bf k}_\perp^2}{\left({\bf k}_\perp^2 + x^2 m^2\right)^2}\, .
     \label{3.2}
\end{equation}
By construction, this is the vacuum term in (\ref{2.4}).
It shows the dead cone effect: gluon radiation is suppressed for
gluons which are emitted under small angles 
\begin{equation}
  \frac{{\bf k}_\perp^2}{\omega^2} < \frac{m^2}{E^2}\, .
  \label{3.3}
\end{equation}
The vacuum term (\ref{3.2}) for the massive case vanishes for 
$k_\perp \to 0$, while the corresponding massless limit formally 
diverges like $\frac{1}{k_\perp^2}$.

\subsection{$N=1$ Opacity correction to the dead cone effect}
\label{sec3b}
%
We consider a medium of spatially extended QCD matter, modeled as 
a collection of colored Yukawa-type scattering centers (\ref{a.6}) 
with Debye screening mass $\mu$. To first order $N=1$ in opacity, 
the average momentum transfer from the medium
to the hard quark is $\mu$. 

{\it Qualitative arguments:} 
Consider a massless quark first.
A gluon of energy $\omega$ decoheres from the wave function of this
quark if its typical formation
time $\bar{t}_{\rm coh} = \frac{2\omega}{\mu^2}$ is smaller than the
typical distance $L$ between the production point of the parton
and the position of the scatterer. Hence, gluon radiation occurs
if the phase 
\begin{equation}
  \bar{\gamma} 
  = \frac{L}{\bar{t}_{\rm coh}} \equiv \frac{\bar{\omega}_c}{\omega}
  \, ,
  \label{3.4}
\end{equation}
exceeds unity. Here,  the characteristic gluon energy is
\begin{equation}
  \bar\omega_c = \frac{1}{2} \mu^2\, L\, .
  \label{3.5}
\end{equation}
To discuss the transverse momentum dependence of gluon emission,
we consider the corresponding ${\bf k}_\perp$-dependent 
phase accumulated due to scattering of the gluon,
\begin{equation}
  \frac{ {\bf k}_\perp^2}{2\omega} L = \frac{ 
  \bar{\kappa}^2}{\omega/ \bar{\omega}_c}
  > 1\, ,
  \label{3.6}
\end{equation}
where we use the rescaled transverse momentum 
\begin{equation}
  \bar{\kappa}^2 = \frac{{\bf k}_\perp^2}{\mu^2}\, .
  \label{3.7} 
\end{equation}
For sufficiently large transverse momentum, the medium-induced
gluon distribution will show the characteristic perturbative
powerlaw for gluon production in a hard process, 
\begin{equation}
  \omega\frac{dI^{\rm med}_{m=0}}{d\omega\, d\bar{\kappa}^2}
  \propto \frac{1}{\bar{\kappa}^4}\, \qquad
  \hbox{for}\qquad \bar{\kappa} > 1\, .
  \label{3.8}
\end{equation}
This behavior can be checked in the $N=1$ opacity expansion by 
taking the large-$\bar{\kappa}$ limit of Eq.~(\ref{3.14}) below. For 
transverse momentum $\bar{\kappa} < 1$, the gluon distribution levels 
off to a constant~\cite{Salgado:2003rv} which depends on 
$\omega/\bar{\omega}_c$. According to the condition (\ref{3.6}),
gluons can be emitted only for 
$\bar{\kappa}^2 > \frac{\omega}{\bar{\omega}_c}$. This allows to 
estimate from (\ref{3.8}) the transverse momentum integrated
gluon distribution in the region $\omega > \bar{\omega}_c$
where only $\bar{\kappa} > 1$ is relevant,
\begin{equation}
    \omega\frac{dI^{\rm med}_{m=0}}{d\omega}
    \propto \int_{\omega/\bar{\omega}_c}^\infty
    \frac{d\bar{\kappa}^2}{\bar{\kappa}^4}
    \propto \frac{\bar{\omega}_c}{\omega}\,  \qquad
    \hbox{for}\qquad \omega > \bar{\omega}_c\, .
    \label{3.9}
\end{equation}
This large-$\omega$ behavior has been established for the $N=1$
opacity approximation~\cite{Gyulassy:2000fs,Salgado:2003gb}.

To estimate the mass-dependence of the gluon distribution,
we first introduce the dead cone factor
\begin{equation}
   F(\bar{\kappa}, \bar{M}) = 
   \left( \frac{\bar{\kappa}^2}{\bar{\kappa}^2 + \bar{M}^2} \right)^2
   =    \left( \frac{{\bf k}_\perp^2}{{\bf k}_\perp^2 + x^2 m^2} \right)^2 \, ,
   \label{3.10} 
\end{equation}
where
\begin{equation}
  \bar{M}^2 = \frac{x^2 m^2}{\mu^2} 
  = \frac{1}{2} \left( \frac{m^2}{E^2} \right) 
    \frac{\bar{R}}{\bar{\gamma}^2}
  \, ,\qquad \bar{R} = \bar{\omega}_c\, L\, .
  \label{3.11}
\end{equation}
For the vacuum term (\ref{3.1}), this factor has the property 
that
\begin{equation}
    \omega\frac{dI^{\rm vac}_{m}}{d\omega\, d\bar{\kappa}^2}
    = F(\bar{\kappa}, \bar{M})\, 
        \omega\frac{dI^{\rm vac}_{m=0}}{d\omega\, d\bar{\kappa}^2}\, .
    \label{3.12}
\end{equation}
If the same estimate holds for the medium-induced part of the
gluon radiation, then the distribution (\ref{3.8}) is depleted
for $\bar{\kappa}^2 < {\rm max}\left[\bar{\gamma},\bar{M}^2\right]$. 
For the large-$\omega$ region, where $\bar{M}^2 > \bar{\gamma}$,
the transverse momentum integrated 
gluon distribution can thus be estimated from (\ref{3.8}),
\begin{equation}
    \omega\frac{dI^{\rm med}_{m}}{d\omega}
    \propto \int_{\bar{M}^2}^\infty
    \frac{d\bar{\kappa}^2}{\bar{\kappa}^4}
    \propto 
    \left( \frac{m^2}{E^2} \right) \bar{R}
    \frac{\bar{\omega}_c^2}{\omega^2}\, \quad 
    \hbox{for}\quad \omega > \bar{\omega}_c/\left( \frac{m^2}{E^2}\bar{R}
    \right)^{1/3} .
    \label{3.13}
\end{equation}
This distribution drops off faster ($\propto \frac{1}{\omega^2}$)
than the corresponding massless term (\ref{3.9}). We thus expect a 
mass-dependent depletion of the medium-induced gluon radiation
at large gluon energy.  We emphasize that the dead cone suppression
is a sufficient but not a necessary condition for this
large-$\omega$ behavior. What is needed is only that mass effects
regulate the singularity $\frac{1}{\bar{\kappa}^4}$ on a scale
$\bar{M}$. Since the phase space for $\bar{\kappa}^2 < \bar{M}^2$
is small, it is not essential whether this regulation occurs by
complete extinction of the radiation e.g. via $F(\bar{\kappa}, \bar{M})$,
or whether the dead cone is filled by a finite non-singular spectrum 
for $\bar{\kappa}^2 < \bar{M}^2$. We shall find that the latter case
is realized. 

{\it Quantitative analysis:}
We have calculated the medium modification of the gluon 
energy distribution (\ref{2.4}) to first order in opacity,
see Appendix~\ref{appa}. In the massless limit, 
our result reduces to the $N=1$ opacity result given in 
Ref.~\cite{Wiedemann:2000za}. For finite in-medium path-length
$L$, it interpolates between the totally coherent
and totally incoherent limiting cases of the gluon radiation
spectrum. In particular, the incoherent limit is an independent
superposition of three distinct radiation terms: the hard radiation 
term (\ref{3.2}) shifted to $H({\bf k} + {\bf q})$ due to elastic 
scattering, a Gunion-Bertsch radiation term associated to gluon
production due to a single scattering center well separated from the
quark production, and a probability conserving third term
(see Appendix~\ref{appa}). In terms of the dimensionless
variables introduced above, the medium-modified gluon
distribution to first order in opacity is
\begin{eqnarray}
  \omega\frac{dI(N=1)}{d\omega\, d\bar{\kappa}^2}
  &=& \frac{\alpha_s\, C_F}{\pi} (2\, n_0\, L)  
    \int_0^\infty d\bar{q}^2
    \frac{ \left( \bar{q}^2 + \bar{M}^2 \right)
           - \frac{1}{\bar{\gamma}} 
             \sin\left[ \bar{\gamma}
                     \left(\bar{q}^2 + \bar{M}^2 \right)\right]}
         {\left(\bar{q}^2 + \bar{M}^2 \right)^2}
    \nonumber \\
   && \times 
    \frac{\bar{q}^2}{\bar{q}^2 + \bar{M}^2}
     \frac{ \left( \bar{\kappa}^2 + \bar{M}^2\right)
            + \left( \bar{\kappa}^2 - \bar{M}^2\right)
              \left( \bar{\kappa}^2 - \bar{q}^2\right)}
          { \left( \bar{\kappa}^2 + \bar{M}^2\right)
            \left[ \left( 1 + \bar{\kappa}^2 + \bar{q}^2\right)^2
                   - 4 \bar{\kappa}^2 \bar{q}^2 \right]^{3/2} }\, .
   \label{3.14}
\end{eqnarray}
The medium-induced gluon energy distribution is obtained by
integrating (\ref{3.14}) over transverse momentum up to the
kinematic boundary $k = \omega$. In terms of the integration
variable $\bar{\kappa}^2$, this corresponds to
\begin{equation}
  \omega\frac{dI(N=1)}{d\omega}
  = \int_0^{\bar{R}/2\bar{\gamma}^2} d\bar{\kappa}^2
  \omega\frac{dI(N=1)}{d\omega\, d\bar{\kappa}^2}\, ,
   \label{3.15}
\end{equation}
where the integration boundary is written in terms of
$\bar{R} = \bar\omega_c\, L$. 
The integral of (\ref{3.15}) over gluon energy defines the
average medium-induced energy loss 
\begin{equation}
  \langle \Delta E_{\rm ind}\rangle 
  = \int_0^E d\omega\,   \omega\frac{dI(N=1)}{d\omega}\, . 
  \label{3.16}
\end{equation}
In the $L \to \infty$ limit, equation (\ref{3.16}) coincides with
the expression given by Djordjevic and Gyulassy~\cite{Djordjevic:2003zk}. 
At finite in-medium pathlength $L$, 
the differences between (\ref{3.16}) and Ref.~\cite{Djordjevic:2003zk}
may be due to the use of a different density distribution
$n(\xi)$ of scattering centers.

\subsection{Numerical results}
\label{sec3c}
{\it Explored parameter space:} The double differential gluon 
distribution (\ref{3.14}) for massless quarks depends on two
parameters only, the characteristic gluon energy $\bar{\omega}_c$
and the typical transverse momentum (Debye screening mass) $\mu$.
Presenting our results in rescaled variables $\bar{\kappa}^2
= {\bf k}_\perp^2/ \mu^2$ and $\bar{\gamma} = \bar{\omega}_c/\omega$,
we explore the unrestricted parameter range for $\bar{\omega}_c$
and $\mu^2$. For the transverse momentum integrated gluon distribution
(\ref{3.15}), the parameter $\bar{R} = \bar{\omega}_c\, L$ enters
via the integration boundary. We motivate our choice of the value
of $\bar{R}$ from a model analysis of high-$p_\perp$ suppressed 
hadroproduction in central Au-Au collisions at RHIC. The order of
magnitude of the suppression is in qualitative agreement with the
parameter choice $\bar{R} = 2000$, $\bar{\omega}_c = 67.5$
GeV, and $n_0\, L = 1$, see Ref.~\cite{Salgado:2003gb}. We thus
choose $\bar{R} = 1000$ and a significantly larger value
$\bar{R} = 40000$ for numerical calculations. The latter can be
viewed as an upper estimate for LHC. If the quark mass is finite,
the double differential gluon distribution also depends on the 
effective mass (\ref{3.11}) which is a function of the ratio $m/E$ 
and of $\bar{R}$. We prefer to present all results in terms of 
$m/E$ and $\bar{R}$ although they appear in (\ref{3.14})
only in one combination.

\begin{figure}[t!]\epsfxsize=10.7cm
\centerline{\epsfbox{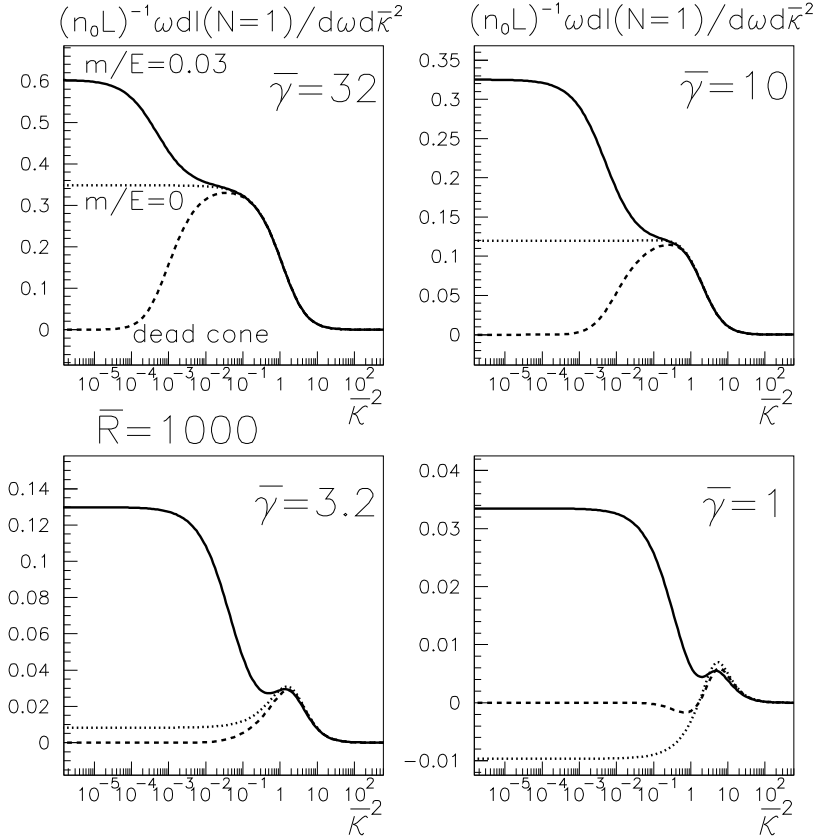}}
\vspace{0.5cm}
\caption{The medium-induced gluon energy distribution as a function
of the transverse momentum $\bar{\kappa}^2 = {\bf k}_\perp^2/\mu^2$ 
and for different values of 
$\bar{\gamma} = \bar{\omega}_c/\omega$. Different curves correspond
to the full medium-induced gluon distribution (\ref{3.14})
for a mass to energy ratio 0.03 of the heavy quark (solid line),
the massless limit (\ref{3.17}) of this spectrum (dotted line), 
and its dead cone approximation (\ref{3.18}) (dashed line). 
}\label{fig1}
\end{figure}

{\it Results:}
We studied the differences between the medium-induced 
gluon distribution (\ref{3.14}) radiated off massive quarks, and the
corresponding double differential gluon distribution  
for massless quarks
\begin{eqnarray}
  \lim_{m \to 0}
  \omega\frac{dI(N=1)}{d\omega\, d\bar{\kappa}^2}
  &=& \frac{\alpha_s\, C_F}{\pi} (2\, n_0\, L)  
    \int_0^\infty d\bar{q}^2
    \frac{ \bar{q}^2 
           - \frac{1}{\bar{\gamma}} 
             \sin\left( \bar{\gamma} \bar{q}^2 \right)}
                     {\bar{q}^4}
    \nonumber \\
   && \qquad \times 
     \frac{ \left( 1 + \bar{\kappa}^2 - \bar{q}^2\right)}
          { \left[ \left( 1 + \bar{\kappa}^2 + \bar{q}^2\right)^2
                   - 4 \bar{\kappa}^2 \bar{q}^2 \right]^{3/2} }\, .
   \label{3.17}
\end{eqnarray}
Also, we tested the conjecture that the medium-induced distribution 
can be obtained from the massless expression by multiplying 
with a dead cone factor (\ref{3.10}),
\begin{eqnarray}
      \omega\frac{dI_{\rm dead}(N=1)}{d\omega\, d\bar{\kappa}^2} &=&
      F(\bar{\kappa}, \bar{M}) \,
      \lim_{m \to 0}
      \omega\frac{dI(N=1)}{d\omega\, d\bar{\kappa}^2}\, .
      \label{3.18}   
\end{eqnarray}
Fig.~\ref{fig1} shows that the transverse momentum dependence of 
the medium-induced gluon distribution deviates qualitatively from the
dead cone approximation (\ref{3.18}). At small transverse momentum,
the medium-induced radiation does not vanish as for the dead cone
approximation (\ref{3.18}). In contrast, it is enhanced compared to
the massless case.  
%

For a more detailed discussion of Fig.~\ref{fig1}, we first note that
the mass to energy ratio $\frac{m}{E}$ and the parameter $\bar{R}$ 
enter the double differential gluon distribution only via the
effective mass parameter $\bar{M}^2  = \frac{1}{2} \frac{m^2}{E^2}
\frac{\bar{R}}{\bar{\gamma}^2}$, see (\ref{3.11}). For the numerical 
values in Fig.~\ref{fig1}, we find $\bar{M}^2 = 0.45/\bar{\gamma}^2$.
In accordance with the qualitative arguments given in Section
~\ref{sec3b}, mass-dependent deviations are seen to become 
significant for transverse momentum $\bar{\kappa}^2 < \bar{M}^2$. 
In particular, for smaller gluon energies $\omega$, i.e. for 
larger values of
$\bar{\gamma} = \frac{\bar{\omega}_c}{\omega}$, the onset of 
mass-dependent deviations is at smaller values of $\bar{\kappa}^2$.
On the other hand, for larger gluon energy (i.e. for small
$\bar{\gamma}$), the transverse momentum distribution of the
radiation spectrum can exceed the Debye screening mass significantly.
The condition $\bar{\kappa}^2 < 1$ provides only a rough estimate for the
upper limit on medium-induced gluon radiation. 
%
\begin{figure}[t!]\epsfxsize=12.7cm
\centerline{\epsfbox{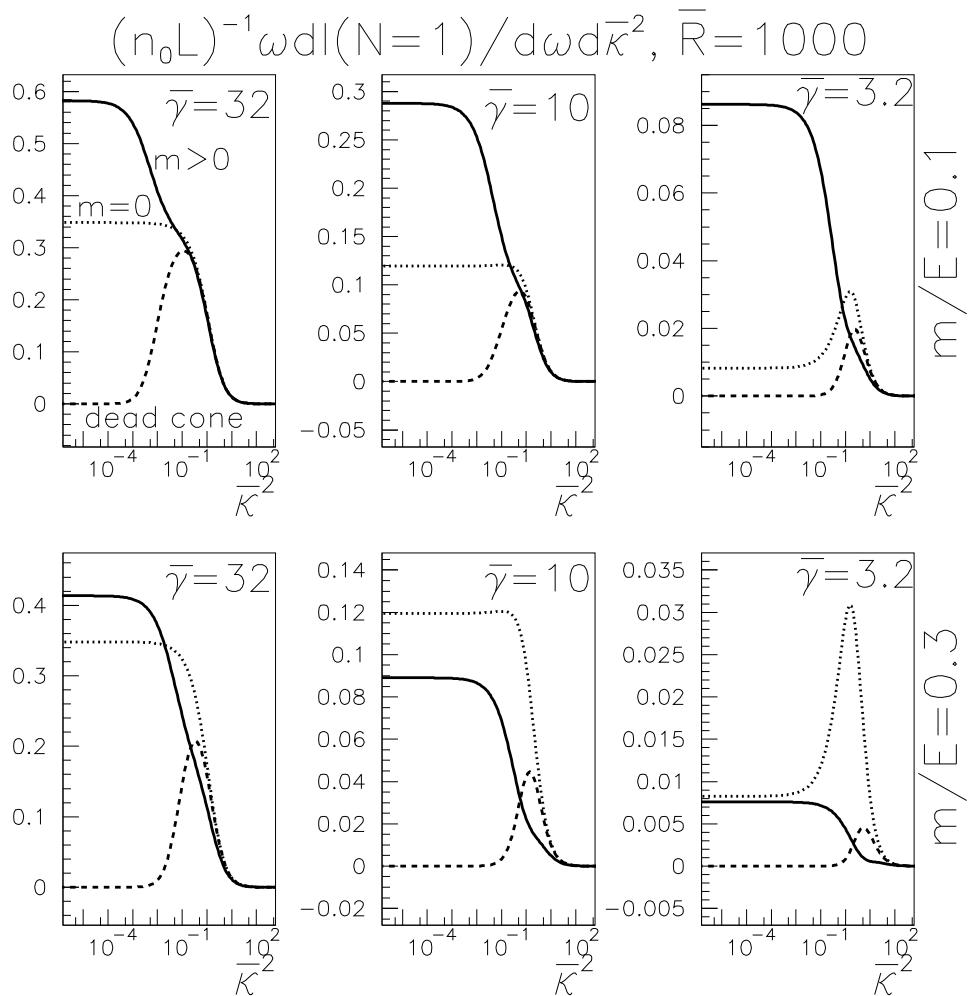}}
\vspace{0.5cm}
\caption{Same as Fig.~\ref{fig1} but for the larger mass to energy
ratios $\frac{m}{E} = 0.1$ and $\frac{m}{E} = 0.3$.
}
\label{fig2}
\end{figure}

We next comment on the origin of the mass dependence in
Fig.~\ref{fig1} which is in qualitative disagreement with
the dead cone prescription (\ref{3.18}).
For massless quarks, medium modifications of
the gluon radiation are known to be determined by two competing
effects~\cite{Wiedemann:2000tf}: First, additional medium-induced 
gluon radiation increases 
the gluon distribution. Second, medium-induced elastic scattering
shifts emitted gluons to larger transverse momentum and thus
depletes the $\propto \frac{1}{k_\perp^2}$ vacuum distribution
at small transverse momentum. [For very energetic gluons which
are emitted predominantly at small angle, the second mechanism
dominates and the medium-induced part of the gluon distribution 
is hence negative for small transverse momentum. This is seen
in the plot for $\bar{\gamma}=1$ in Fig.~\ref{fig1}.] For massive
quarks, the dead cone effect implies that there is no 
vacuum distribution at small angle which can be depleted due to
elastic scattering. As a consequence, for massive quarks the 
second mechanism does not apply at small $\bar{\kappa}$ and the gluon 
radiation is further enhanced.

\begin{figure}[h]
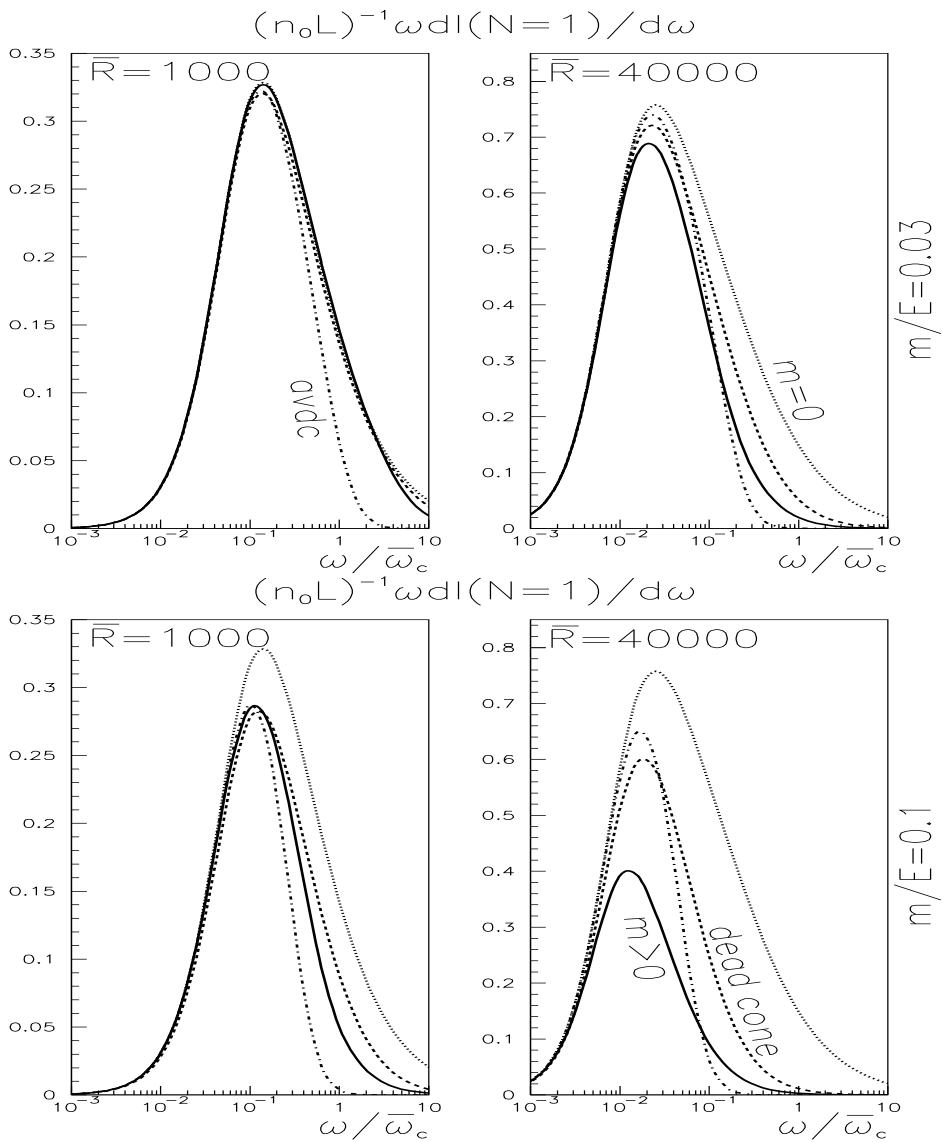

\epsfysize=7.5cm
\epsfxsize=12.4cm
\centerline{\epsfbox{fig3up.epsi}}
%
\epsfysize=7.5cm
\epsfxsize=12.4cm
\centerline{\epsfbox{fig3down.epsi}}
\vspace{0.5cm}
\caption{The medium-induced gluon energy distribution (\ref{3.15})
calculated from the full expression (\ref{3.14}) for a massive
quark (solid line), from the massless limit (\ref{3.17}) (dotted line), 
from the 
dead-cone approximation (\ref{3.18}) (dashed line) and from the 
average dead cone factor (\ref{3.19}) multiplied by the massless 
spectrum (dash-dotted line).  
}\label{fig3}
\end{figure}

We have varied the mass to energy ratio $\frac{m}{E}$ and the
parameter $\bar{R}$ in our calculation over a wide parameter
range ($0.001 < \frac{m}{E} < 0.3$ and $1000 < \bar{R} < 40000$).
Quantitatively, the medium-induced radiation varies
significantly with these parameters (see the discussion below).
Qualitatively, the effect discussed above
is generic for the entire parameter space: For transverse momentum 
$\bar{\kappa} < \bar{M}$, medium-induced gluon radiation fills the dead 
cone. For a significant part of the parameter space, 
it is enhanced compared to 
the massless case, too. For very large mass to energy ratios,
however, mass effects limit significantly the medium-induced
gluon radiation and the radiation from massless quarks finally
dominates for  $\bar{\kappa} < \bar{M}$, see Fig.~\ref{fig2}.

The phase space of the dead cone region $\bar{\kappa}^2 < \bar{M}^2$
is small. For massive quarks, neither the vacuum 
contribution nor the medium-induced contribution to (\ref{2.4})
contain collinear singularities which could enhance the importance of
this phase space region. As a consequence, parton energy loss off
hard quarks will be dominated mainly by radiation in the
region $\bar{\kappa}^2 > \bar{M}^2$. For large $\bar{\kappa}$, 
however, the medium-induced radiation off massive quarks (\ref{3.14}) 
is suppressed compared to the massless limit if the
$\frac{m}{E}$-ratio is sufficiently large, see Fig.~\ref{fig2}.
This feature dominates the transverse momentum integrated gluon 
energy distribution (\ref{3.15}), see Fig.~\ref{fig3}. In
particular, the radiation is depleted for large gluon energy,
since it increases with the effective mass parameter
$\bar{M}^2 = \frac{1}{2} \frac{m^2}{E^2} 
\bar{R} \frac{\omega^2}{\bar{\omega}_c^2}$. Fig.~\ref{fig3}
indicates that 
the mass-dependent suppression of (\ref{3.15}) is even stronger 
than predicted by the dead cone approximation (\ref{3.18}). 

One may ask whether the correct amount of mass-dependent 
suppression can be estimated from a $\bar{\kappa}$-independent
suppression factor $F_{\rm avdc}(\bar{M})$ multiplying the gluon
energy distribution for massless quarks. [Here, the subscript
``{\rm avdc}'' stands for ``{\rm average dead cone}''.] Such a factor 
could be useful for simplified calculations in which 
${\bf k}_\perp$-dependent information is not available. 
Paralleling an estimate given in Ref.~\cite{Dokshitzer:2001zm}
for the dipole approximation [see Eq.~(\ref{4.16}) below],
we have estimated $F_{\rm avdc}(\bar{M})$ by evaluating the 
dead cone factor (\ref{3.7}) 
for the characteristic angle under which medium-induced gluon
radiation occurs, $\theta_c^2 = \frac{\mu^2}{\omega^2} =
\left( \bar{\omega}_c \over \omega \right)^2 \frac{1}{\bar{R}}$.
We find
\begin{equation}
  F_{\rm avdc}(\bar{M}) = 
  \left( \frac{1}{ 1 + \bar{M}^2} \right)^2 \, .
  \label{3.19}
\end{equation}
Substituting this average for the dead-cone factor in (\ref{3.18})
and calculating the transverse momentum integrated spectrum, one
tends to overestimate the mass-dependent reduction of the 
medium-induced radiation, see Fig.~\ref{fig3}.

\begin{figure}[h]
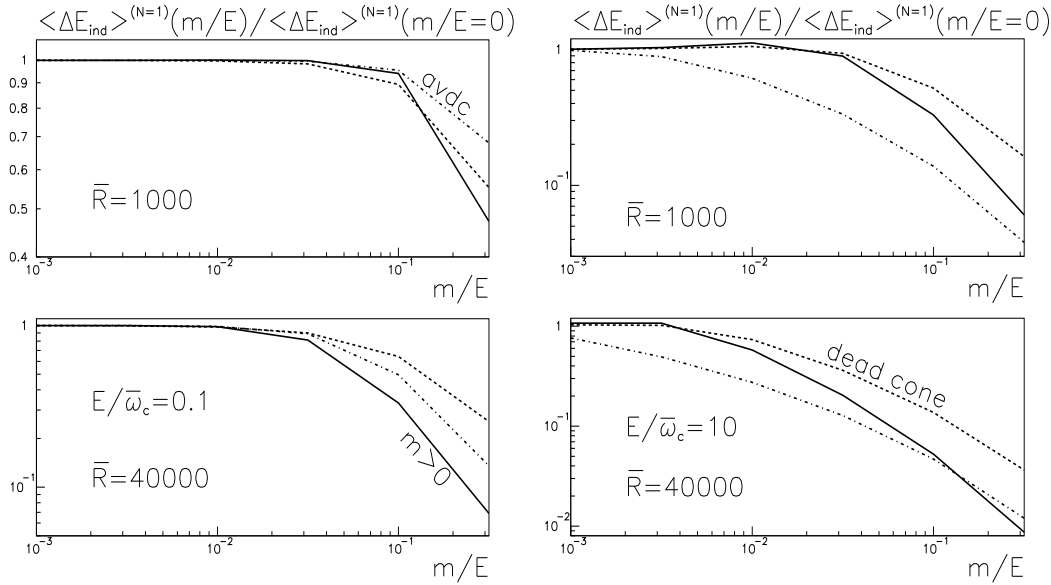

\begin{minipage}[t]{68mm}
\epsfxsize=6.7cm
\epsfysize=7.7cm
\centerline{\epsfbox{fig4left.epsi}}
\end{minipage}
\hspace{\fill}
\begin{minipage}[t]{68mm}
\epsfxsize=6.7cm
\epsfysize=7.7cm
\centerline{\epsfbox{fig4right.epsi}}
\end{minipage}
\vspace{0.5cm}
\caption{The average medium-induced parton energy loss (\ref{3.16})
for massive quarks, normalized to the massless limit, for different
values of the density parameter $\bar{R} = \bar{\omega}_c\, L$.
Curves are calculated for the full medium-induced radiation
(\ref{3.14}) off massive quarks (solid lines), the dead cone
approximation (\ref{3.18}) (dashed lines) and the corresponding
expression with averaged dead cone (\ref{3.19}) (dash-dotted lines).
}\label{fig4}
\end{figure}

Fig.~\ref{fig4} shows the ratio of the medium-induced average
parton energy loss (\ref{3.16}) for massive quarks, normalized
to the same quantity calculated for massless quarks. In general,
a finite quark mass is found to reduce the parton energy loss
and this reduction increases strongly with $\frac{m}{E}$.

\begin{figure}[h]
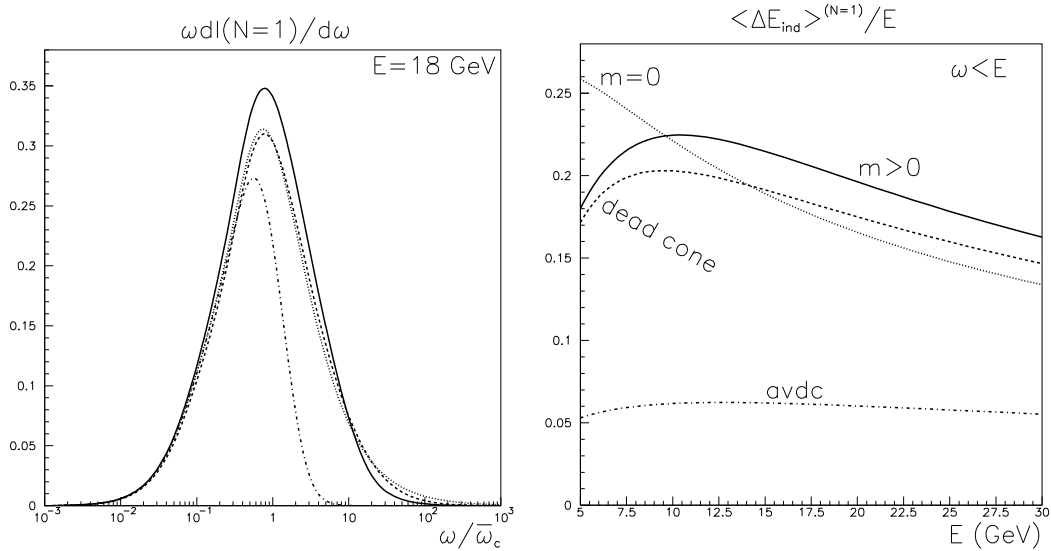

\begin{minipage}[t]{68mm}
\epsfxsize=6.7cm
\centerline{\epsfbox{fig5left.epsi}}
\end{minipage}
\hspace{\fill}
\begin{minipage}[t]{68mm}
\epsfxsize=6.7cm
\centerline{\epsfbox{fig5right.epsi}}
\end{minipage}
\vspace{0.5cm}
\caption{For finite quark energy $E$, the 
normalized average energy loss (rhs) depends significantly on the 
kinematic boundary up to which the gluon energy distribution
(lhs) is integrated. This entails significant uncertainties,
which are discussed in the text. Parameter values are taken from
Ref.~\cite{Djordjevic:2003zk}.
}\label{fig5}
\end{figure}

We now discuss in more detail the main theoretical uncertainty 
entering (\ref{3.16}) and Fig.~\ref{fig4}. Since the double 
differential distribution (\ref{3.14}) is calculated in the 
eikonal approximation, it can have support for gluon energies 
$\omega > E$. This just indicates the limited validity of
the eikonal approximation for small energies $E$. This issue
is critical for calculations of the average parton energy loss 
(\ref{3.16}) which can depend strongly on the phase space limit 
$\frac{\omega}{\bar{\omega}_c} < \frac{E}{\bar{\omega}_c}$ 
imposed on the integral over the gluon energy distribution.
In Fig.~\ref{fig5}, we illustrate this point by showing the
gluon energy distribution and the average energy loss (\ref{3.16}) 
for the parameter set $n_0\, L = 4$, $\mu = 500$ MeV, $L = 4$
fm and $\alpha_s = 0.3$. This parameter set was used to reproduce 
the suppression of high-$p_\perp$ hadroproduction observed in Au+Au 
collisions at RHIC~\cite{Levai:2001dc}, and it is used in the 
numerical calculation of Ref.~\cite{Djordjevic:2003zk} together
with a charm quark mass of $m= 1.5$ GeV. 
For the standard upper integration 
bound $\omega < E$, the massive case agrees qualitatively in magnitude 
and energy dependence with the calculation shown in Fig. 1 of
Ref.~\cite{Djordjevic:2003zk}. Small quantitative differences 
are due to the different finite $L$ dependence of the density 
distributions, as argued following Eq. (\ref{3.16}). However,
as seen in Fig.~\ref{fig5}, the average energy loss (\ref{3.16}) 
off massless quarks turns out to be smaller since a larger part 
of the gluon energy distribution lies above the kinematic cut.
We have made no attempt to remedy this possibly unphysical behavior.
We simply emphasize that any {\it a posteriori} modification of
the large-$\omega$ tail of $\omega \frac{dI}{d\omega}$ entails
significant theoretical uncertainties. It is an open problem of
obvious importance to include finite energy constraints on the
level of the radiation spectrum (i.e. on the level of multiple
scattering Feynman diagrams) rather than on the
level of the integrated parton energy loss.

\section{Dipole approximation (multiple soft scattering)}
\label{sec4}
If the medium provides a large number of soft momentum transfers,
rather than a few harder ones, then the projectile performs a 
Brownian motion in transverse momentum. This 
case can be studied in the saddle point approximation of the
path-integral (\ref{2.1}), using~\cite{Zakharov:1996fv,Zakharov:1998sv}
\begin{eqnarray}
  n(\xi)\, \sigma({\bf r}) \simeq \frac{1}{2}\, \hat{q}(\xi)\, {\bf r}^2\, .
  \label{4.1}
\end{eqnarray}
Here, $\hat{q}(\xi)$ is the transport coefficient\cite{Baier:1996sk} 
which characterizes the medium-induced transverse momentum squared 
$\langle q_\perp^2\rangle$ transferred to the projectile per unit 
path length $\lambda$. In the approximation (\ref{4.1}), the path 
integral in (\ref{2.1}) is equivalent to that of a harmonic oscillator
which allows for an explicit calculation. 

\subsection{The Vacuum Term shows the Dead Cone Effect}
\label{sec4a}
For the transport coefficient $\hat{q}(\xi) = \hat{q} \Theta(L-\xi)$
of a static medium of length $L$, we evaluate the gluon
distribution (\ref{2.1}) by splitting the longitudinal
integrals into three parts~\cite{Wiedemann:1999fq},
\begin{eqnarray}
  \omega \frac{dI}{d\omega\, d\kappa^2} &=& 
  \omega \frac{dI_4}{d\omega\, d\kappa^2} + 
  \omega \frac{dI_5}{d\omega\, d\kappa^2} + 
  \omega \frac{dI_6}{d\omega\, d\kappa^2} 
  \nonumber\\
  &=& \int_0^L dy_l \int_{y_l}^L d\bar{y}_l  \dots
  + \int_0^L dy_l \int_L^{\infty} d\bar{y}_l  \dots
  + \int_L^{\infty} dy_l \int_{y_l}^{\infty} d\bar{y}_l  \dots
  \label{4.2}
\end{eqnarray}
In complete analogy to the calculation for the massless 
case~\cite{Wiedemann:2000tf},
one can shown that the term $I_6$ does not depend on the 
medium and takes the form
\begin{equation}
  \omega \frac{dI_6}{d\omega\, d{\bf k}_\perp} =
  \frac{\alpha_s\, C_F}{\pi^2}
  \frac{{\bf k}_\perp^2}{\left({\bf k}_\perp^2 + x^2 m^2\right)^2}\, .
  \label{4.3}
\end{equation}
This expression coincides with the vacuum term (\ref{3.1}). 
In the multiple soft scattering separation, there is hence
a simple separation of the gluon distribution (\ref{2.4}) into 
the vacuum term (\ref{4.3}) and the medium-induced contribution
$I_4 + I_5$. The latter vanish in the absence of a medium,
$\hat{q} = 0$. 

\subsection{Medium-induced gluon radiation}
\label{sec4b}
We now discuss the medium-induced part of the gluon distribution 
(\ref{2.4}) which in the multiple soft scattering (mss) approximation
takes the form
\begin{eqnarray}
  \omega \frac{dI^{\rm mss}}{d\omega\, d\kappa^2} &=& 
  \omega \frac{dI_4}{d\omega\, d\kappa^2} + 
  \omega \frac{dI_5}{d\omega\, d\kappa^2} \, .
  \label{4.4} 
\end{eqnarray}
The terms $I_4$ and $I_5$ are defined in (\ref{4.2}) and are
given explicitly in Appendix~\ref{appb}. 

{\it Qualitative arguments:}
Consider first the qualitative behavior of (\ref{4.4}) for
the case of a massless quark.
The energy and transverse momentum scales which determine 
medium-induced gluon radiation can be estimated in analogy to 
the discussion given for the $N=1$ opacity case. The phase
accumulated by the gluon due to multiple scattering is given
in terms of the characteristic gluon energy $\omega_c$~\cite{Baier:2002tc},
\begin{equation}
  \gamma = \Bigg\langle \frac{k_\perp^2}{2\omega}\, \Delta z \Bigg\rangle
  \sim \frac{\hat{q}\, L}{2\omega} L = \frac{\omega_c}{\omega}\, ,
  \qquad \omega_c = \frac{1}{2}\, \hat{q}\, L^2\, .
  \label{4.5}
\end{equation}
This phase should be larger than $\gamma \sim O(1)$ for the 
gluon to decohere from the massless quark. Here, we have used
that the gluon carries a typical transverse momentum squared
of order $\langle {\bf k}_\perp^2 \rangle \sim \hat{q}\, L$. 
We express the transverse momentum ${\bf k}_\perp$ in units of
this characteristic scale,
\begin{equation}
  \kappa^2 = \frac{{\bf k}_\perp^2}{\hat{q}\, L}\, .
  \label{4.6}
\end{equation}
For the case of many soft scattering centers, several centers 
act coherently if the formation time of the gluon exceeds its
mean free path. The typical transverse momentum accumulated by
the gluon is then ${\bf k}_\perp^2 \simeq \langle q_\perp^2 \rangle\, 
\frac{t_{\rm form}}{\lambda} = \hat{q}\, t_{\rm form}$ while its 
formation time is
$t_{\rm form} \simeq \frac{\omega}{{\bf k}_\perp^2}$. 
Medium-induced radiation thus relates transverse momentum 
and gluon energy,
\begin{equation}
  {\bf k}_\perp^2 \simeq \sqrt{\hat{q}\, \omega}
  \qquad \hbox{or in dimensionless units:}
  \qquad  \kappa^2 \simeq \sqrt{\frac{1}{2\, \gamma}}\, . 
  \label{4.7}
\end{equation}
On the other hand, the number of scattering centers which 
act coherently is $N_{\rm coh} = \frac{t_{\rm form}}{\lambda}$.
For $N_{\rm coh} > 1$, the medium-induced gluon radiation
in the multiple soft scattering limit reads
\begin{equation}
  \omega\frac{dI^{\rm mss}_{m=0}}{d\omega\, d\kappa^2} \simeq
    \frac{L}{\lambda\, N_{\rm coh}}\, 
  \omega\frac{dI_{m=0}^{1{\rm scatt}}}{d\omega\, d\kappa^2}\, ,
  \label{4.8}
\end{equation}
where $I_{m=0}^{1{\rm scatt}}$ is the radiation off a single scattering
center and the total number $L/\lambda$ of scattering centers along
the path of extension $L$ is reduced by coherence effects to the
effective number
\begin{equation}
  \frac{L}{\lambda\, N_{\rm coh}} \sim \sqrt{\frac{\omega_c}{\omega}}
  \sim \frac{1}{\kappa^2} \, .
  \label{4.9}
\end{equation}
If the gluon energy is small, $\omega < \omega_c$, then it follows 
from (\ref{4.7}) that $\kappa^2 < 1$. The transverse momentum
integral over $\omega\frac{dI_{m=0}^{1{\rm scatt}}}{d\omega\, d\kappa^2}$
gives an energy-independent function and we find from (\ref{4.8})
and (\ref{4.9}) that
\begin{equation}
  \omega\frac{dI^{\rm mss}_{m=0}}{d\omega} \propto
    \frac{L}{\lambda\, N_{\rm coh}}\, 
    \sim \sqrt{\frac{\omega_c}{\omega}}\, \qquad
    \hbox{for}\qquad \omega < \omega_c\, .
  \label{4.10}
\end{equation}
For $\kappa^2 > 1$, on the other hand, the spectrum
$\omega\frac{dI_{m=0}^{1{\rm scatt}}}{d\omega\, d\kappa^2}$ is characterized
again by the perturbative tail (\ref{3.8}). Combining this 
information with Eqs. (\ref{4.8}) and (\ref{4.9}), one
obtains
\begin{equation}
  \omega\frac{dI^{\rm mss}_{m=0}}{d\omega} \sim
  \int_{\omega/\omega_c}^{\infty}
  \frac{d\kappa^2}{\kappa^6}
  \sim \left( \frac{\omega_c}{\omega}\right)^2\, \qquad
    \hbox{for}\qquad \omega > \omega_c\, .
  \label{4.11}
\end{equation}
Both limiting cases, (\ref{4.10}) and (\ref{4.11}) agree with the
results of the full calculation~\cite{Salgado:2003gb} in the
multiple soft scattering limit for massless quarks.

To estimate the mass-dependence of the gluon distribution, 
we parallel the discussion of Section~\ref{sec3b}. We introduce the
dead cone factor
\begin{equation}
   F(\kappa, M) = 
   \left( \frac{\kappa^2}{\kappa^2 + M^2} \right)^2
   =    \left( \frac{{\bf k}_\perp^2}{{\bf k}_\perp^2 + x^2 m^2} \right)^2 \, ,
   \label{4.12} 
\end{equation}
where
\begin{equation}
  M^2 = \frac{x^2 m^2}{\hat{q} L} 
  = \frac{1}{2} \left( \frac{m^2}{E^2} \right)\, R\, 
    \frac{\omega^2}{\omega_c^2}
  \, ,\qquad R = \omega_c\, L\, .
  \label{4.13}
\end{equation}
Restricting the $\kappa^2$-integration of the perturbative tail
by this mass term, we find
\begin{equation}
  \omega\frac{dI^{\rm mss}_{m}}{d\omega} \sim
  \int_{M^2}^{\infty}
  \frac{d\kappa^2}{\kappa^6}
  \sim \frac{1}{M^4} \, ,\qquad
    \hbox{for}\qquad \omega > \omega_c\, .
  \label{4.14}
\end{equation}
In close analogy to the estimate for single hard scattering in
Section~\ref{sec3}, the large-$\omega$ region drops off significantly
faster than for the massless case (\ref{4.11}). We thus expect
a mass-dependent depletion for large $\omega$. 

\subsection{Numerical results}
\label{sec4c}
{\it Explored parameter space:} In analogy to the study in 
Section~\ref{sec3c}, we present all results in rescaled 
variables $\kappa^2$ and $\gamma$. The parameter $R=\omega_c\, L$
is explored for the values 1000 and 40000. See Section~\ref{sec3c}
for further details.

\begin{figure}[h]\epsfxsize=10.7cm
\centerline{\epsfbox{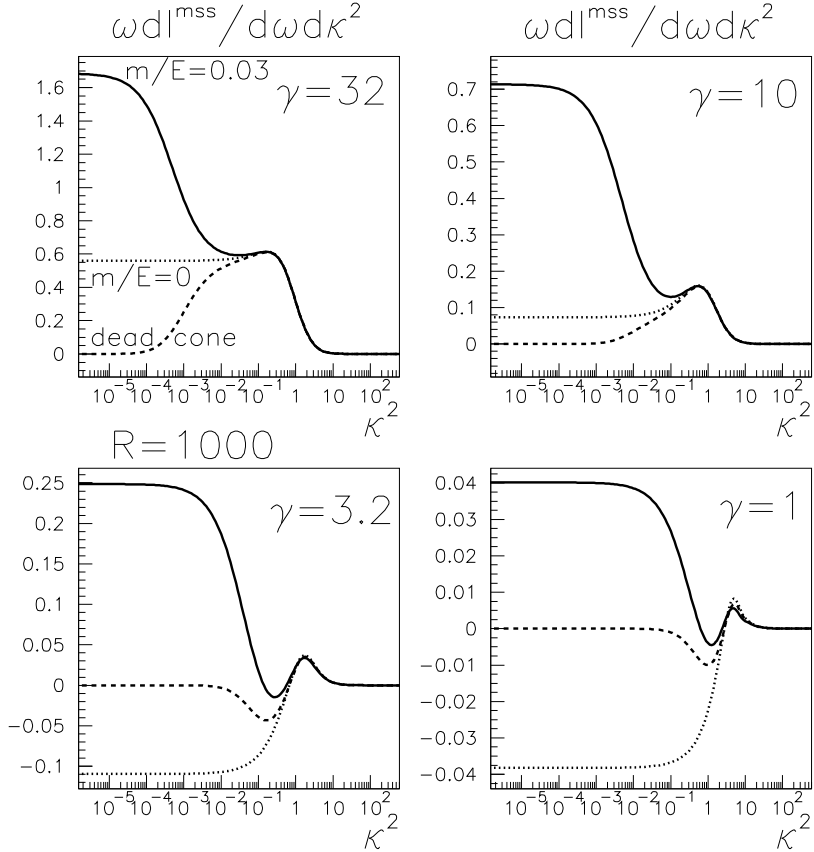}}
\vspace{0.5cm}
\caption{
The medium-induced gluon energy distribution as a function
of the transverse momentum $\kappa^2 = \frac{{\bf k}^2_\perp}{\hat{q}L}$ 
and for different values of 
$\gamma = {\omega_c}/\omega$, calculated in the multiple soft
scattering approximation. Different curves correspond
to the full medium-induced gluon distribution 
for a mass to energy ratio 0.03 of the heavy quark (solid line),
the massless limit of this spectrum (dotted line), 
and its dead cone approximation (\ref{4.15}) (dashed line). 
}\label{fig6}
\end{figure}

{\it Results:} 
We have calculated the transverse momentum dependence 
of the medium-induced gluon distribution (\ref{4.4}) in the
multiple soft scattering approximation. We have compared this
distribution to the massless limit and to the dead cone approximation
\begin{eqnarray}
      \omega\frac{dI^{\rm mss}_{\rm dead}}{d\omega\, d\kappa^2} &=&
      F(\kappa, M) \, 
          \omega\frac{dI^{\rm mss}_{m=0}}{d\omega\, d\kappa^2}\, .
      \label{4.15}   
\end{eqnarray}
As seen in Fig.~\ref{fig6},
the results are in qualitative agreement with those obtained 
in the opacity expansion (see Fig.~\ref{fig3} and Section~\ref{sec3c}
for discussion). In particular, we find that the dead cone is
filled by medium-induced gluon radiation. This feature persists
to larger mass to energy ratios (data not shown) in qualitative 
agreement with the opacity expansion in Fig.~\ref{fig2}.
%

\begin{figure}[h]
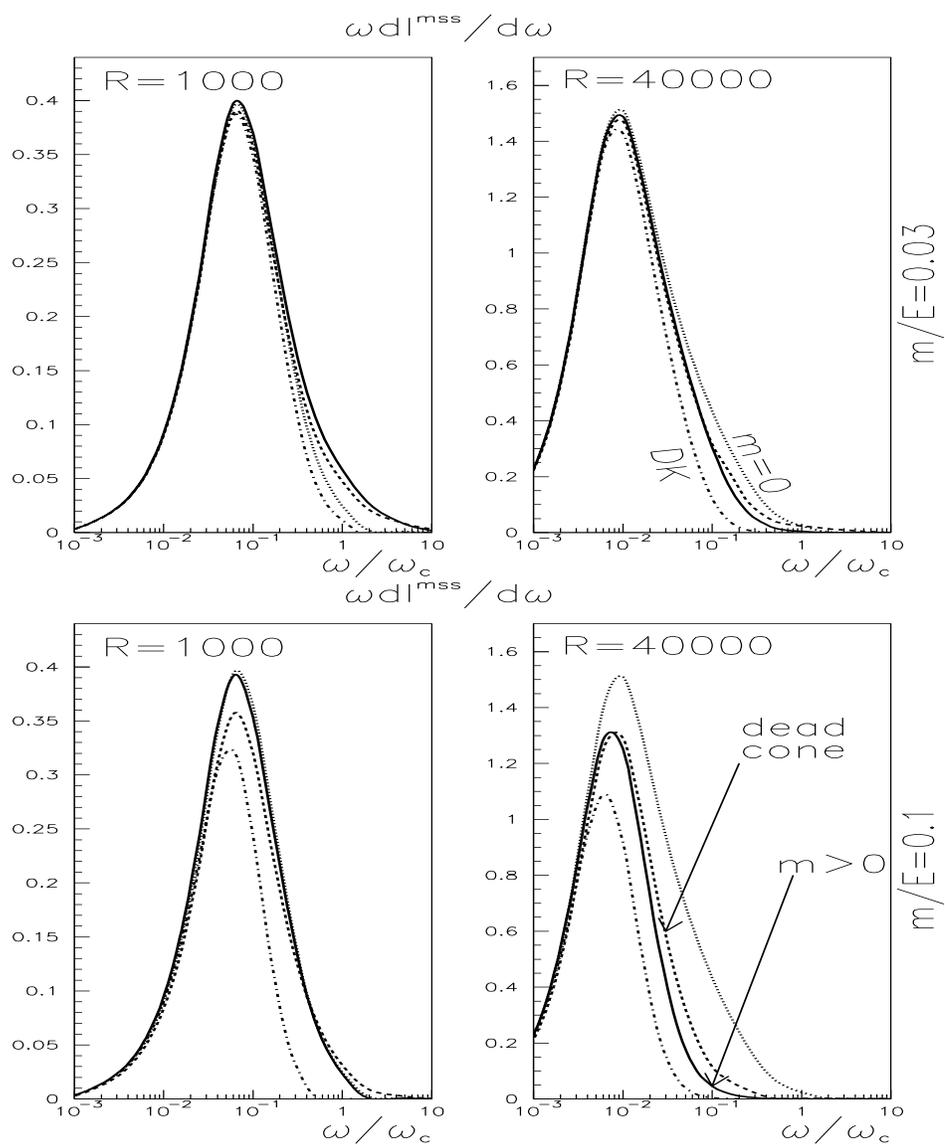

\epsfysize=7.5cm
\epsfxsize=12.4cm
\centerline{\epsfbox{fig7up.epsi}}
\epsfysize=7.5cm
\epsfxsize=12.4cm
\centerline{\epsfbox{fig7down.epsi}}
\vspace{0.5cm}
\caption{Same as Fig.~\ref{fig3}, here calculated in the multiple
soft scattering approximation.
}\label{fig7}
\end{figure}

The transverse momentum integrated gluon energy distribution
calculated from (\ref{4.4}) is shown in Fig.~\ref{fig7}. 
This integral is dominated by the phase space region $\kappa^2 > M^2$,
and thus does not depend strongly on the finite medium-induced radiation 
inside the dead cone region. Compared to the massless case, 
the large-$\omega$ tail of the gluon energy distribution is 
depleted with increasing $m/E$-ratio. The approximation of this 
effect by the dead cone approximation (\ref{4.15}) tends to 
underestimate this depletion. These
findings are in qualitative agreement with the results reported
for the opacity expansion in Fig.~\ref{fig3} and with the qualitative 
expectations based on (\ref{4.14}). 

We have also tested numerically the approximation of Dokshitzer and 
Kharzeev~\cite{Dokshitzer:2001zm}. These authors replaced the 
transverse momentum dependent dead cone factor in
(\ref{4.15}) by an average expression, evaluated at the
characteristic angle $\theta_c^2 \simeq \sqrt{2} \frac{\gamma^{3/2}}{R}$
under which medium-induced gluon radiation occurs on average,
\begin{equation}
  F_{\rm DK} = \left(  \frac{1}{
            1 + \frac{1}{\sqrt{2}} \frac{m^2}{E^2} \frac{R}{\gamma^{3/2}}} 
                \right)^2\, .
  \label{4.16}
\end{equation}
This approximation allows to mimic the qualitative trend but 
is found to overestimate the depletion of the large-$\omega$ 
tail significantly, see Fig.~\ref{fig7}.

\begin{figure}[h]
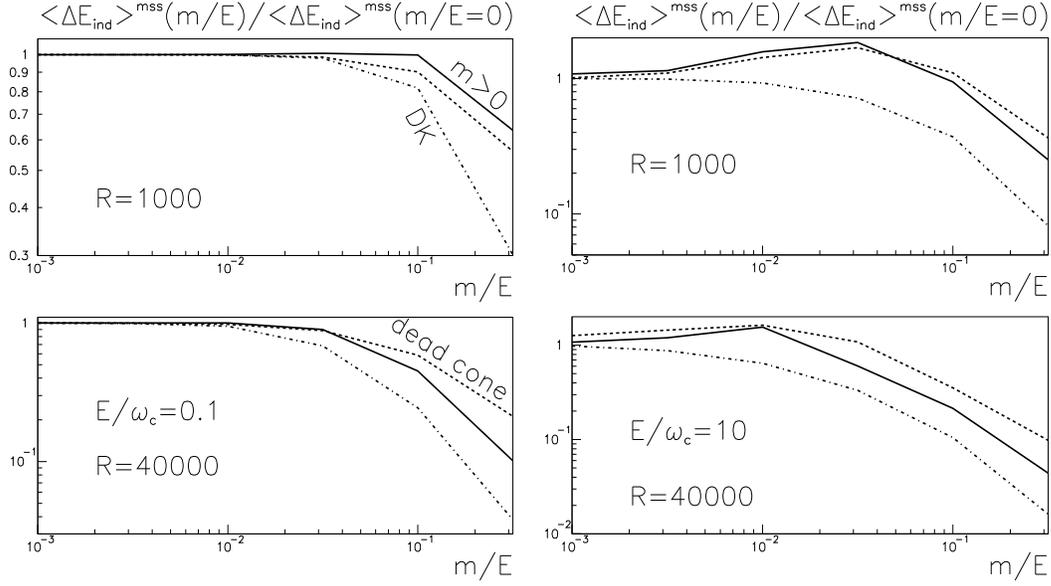

\begin{minipage}[t]{68mm}
\epsfxsize=6.7cm
\epsfysize=7.7cm
\centerline{\epsfbox{fig8left.epsi}}
\end{minipage}
\hspace{\fill}
\begin{minipage}[t]{68mm}
\epsfxsize=6.7cm
\epsfysize=7.7cm
\centerline{\epsfbox{fig8right.epsi}}
\end{minipage}
\vspace{0.5cm}
\caption{Same as Fig.~\ref{fig4}, here calculated in the multiple
soft scattering approximation.
}\label{fig8}
\end{figure}

\begin{figure}[h]
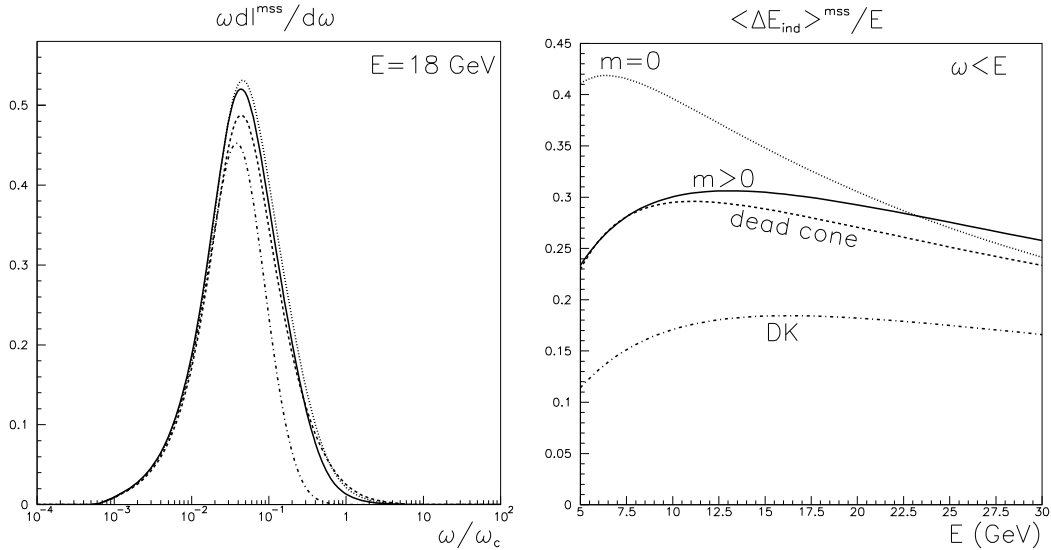

\begin{minipage}[t]{68mm}
\epsfxsize=6.7cm
\centerline{\epsfbox{fig9left.epsi}}
\end{minipage}
\hspace{\fill}
\begin{minipage}[t]{68mm}
\epsfxsize=6.7cm
\centerline{\epsfbox{fig9right.epsi}}
\end{minipage}
\vspace{0.5cm}
\caption{Same as Fig.~\ref{fig5}, here calculated in the multiple
soft scattering approximation for parameter values $R= 2000$,
$\omega_c = 67.5$ GeV and $m = 1.5$ GeV. This figure allows to
illustrate the uncertainties in calculations of the average parton
energy loss for finite energy quarks, see text.
}\label{fig9}
\end{figure}

From the gluon energy distribution, we have calculated the average 
parton energy loss according to (\ref{3.16}). As seen in 
Fig.~\ref{fig8}, a finite quark mass reduces parton 
energy loss significantly for sufficiently large mass to
energy ratios, $m/E > 0.1$ say. For smaller mass to energy ratios,
there is some parameter range where our numerical results indicate
the opposite effect. However, similar to the case of the opacity expansion,
this may be an artifact of the phase space constraint in the definition of
the average energy loss (\ref{3.16}) [see the discussion of 
Figs.~\ref{fig4} and Fig.~\ref{fig5} above]. To explore this
theoretical uncertainty, we have paralleled the logic of Section
~\ref{sec3}. We have calculated in Fig.~\ref{fig9} the gluon energy 
distribution (\ref{4.4}) for the set of parameters which reproduce
the nuclear modification factor for central Au+Au collisions at
RHIC~\cite{Salgado:2003gb}. Integrating this gluon energy distribution
for $\omega < E$, we find again that the average parton energy loss
for massless quarks can be larger than that for massive ones, simply
because the kinematic boundary $\omega < E$ cuts off the more
pronounced large $\omega$-tail of the gluon energy distribution
for $m=0$. For the reasons given in Section~\ref{sec3c}, we conclude
that this motivates to go beyond the high-energy approximation and to
include finite energy constraints in the calculation of 
$\omega \frac{dI}{d\omega}$ rather than to impose them {\it a posteriori}
in the integral over $\omega \frac{dI}{d\omega}$.

\section{Conclusion}
\label{sec5}

Early studies demonstrated that  
the transverse momentum spectra of charmed and beauty hadrons in 
nucleus-nucleus collisions depend strongly on the assumed parton energy 
loss~\cite{Shuryak:1996gc,Lin:1997cn,Mustafa:1997pm}. More recently,
the quark mass dependence of this effect was argued to give access to 
the detailed mechanism of medium-modified parton fragmentation~\cite{Dokshitzer:2001zm,Djordjevic:2003qk,Djordjevic:2003be,Djordjevic:2003zk,Zhang:2003wk}. 
To better assess this mechanism,
we have presented here the first study of the transverse momentum 
and energy dependence of the medium-induced gluon radiation off 
massive quarks. 

Our calculation is based on the path integral formalism (\ref{2.1})
which resums medium-modifications to leading order in $1/E$ and to
all orders in opacity. We have employed two approximation 
schemes which model the medium-dependence of the hard parton fragmentation
as a series of many soft gluon exchanges or as a single semi-hard momentum
transfer. Despite these different physical pictures, both approaches 
lead to comparable results, in agreement with an earlier study of the
massless case~\cite{Salgado:2003gb}.   

We find that medium-induced gluon radiation generically fills the 
dead cone $\theta < m/E$ [see Figs.~\ref{fig1}, ~\ref{fig2} and
~\ref{fig6}]. However, in comparison to the transverse 
momentum integrated gluon energy distribution radiated off massless 
quarks, the radiation is depleted at large gluon energies 
[see Figs.~\ref{fig3} and ~\ref{fig7}].
The average parton energy loss results from an interplay of
both effects and tends to be smaller for massive quarks than 
for massless ones [see Figs.~\ref{fig4} and ~\ref{fig8}]. However, 
for sufficiently small parton energies $E$, the kinematic boundary 
$\omega < E$ can lead to the peculiar case that the average parton 
energy loss is larger for massive quarks than for massless ones 
[see Figs.~\ref{fig5} and ~\ref{fig9}]. We have
argued that this behavior may be unphysical and 
limits the application of the formalism of 
Refs.~\cite{Baier:1996sk,Zakharov:1997uu,Wiedemann:2000za,Gyulassy:2000er,Wang:2001if} to sufficiently large quark energies, see discussion at the
end of Section~\ref{sec3c}.

As of today, the only experimental information about open charm 
production in Au+Au collisions at RHIC is the prompt electron 
spectrum measured by PHENIX at 
$\sqrt{s_{\rm NN}} = 130$ GeV ~\cite{Adcox:2002cg}. It was argued 
that these data do not indicate a significant parton energy loss 
for charm quarks. But it is equally true that they do not 
constrain parton energy loss significantly: First, the experimental 
errors on the prompt electron spectrum are still large. Second,
for the measured transverse momentum range 
$p_\perp^{\rm electron} < 3$ GeV, the correlation between the transverse 
momentum of the electron and of the charmed hadron is very weak. Both 
complicate any conclusion about the medium-induced degradation 
of $p_\perp^{\rm charm}$. In addition, the moderate values 
of $p_\perp^{\rm charm}$ accessed by PHENIX correspond to charm quarks 
which move rather slowly and thus may turn into hadrons
prior to leaving the collision region. In this case, the energy
degradation of charmed hadrons would get contributions from their 
hadronic cross sections and the formalism employed here has to
be revisited.

The experimental information from RHIC is expected to improve soon. 
First, the higher statistics of future runs will allow to measure
the prompt electron spectrum in a wider $p_\perp$ range. Second, 
the topological reconstruction of the hadronic
decay of charmed hadrons ($D^0 \to K^- \pi^+$) should provide a more
direct measurement out to significant transverse momentum. On
a longer time scale, open charm measurements at LHC will further extend
this $p_\perp$ range to $p_\perp^{D^0} \sim 15$ GeV~\cite{Dainese:2003wq}.  
Also, the energy loss of beauty quarks is expected
to become accessible at LHC via high-mass dimuon 
and secondary  J/$\Psi$ production~\cite{Lokhtin:ay}.
Despite the significant uncertainties of our calculation discussed 
above, there is one conclusion which we can draw: parton energy loss
is reduced by mass effects, but for realistic parameter values it
remains sizable. Thus, our study favors a medium-induced 
enhancement of the $D^0/\pi^0$ ratio at sufficiently large transverse 
energy but we still expect the nuclear modification factor for $D^0$ 
to lie below unity.

{\bf Acknowledgment:} We thank R. Baier, A. Dainese, K. Eskola,
H. Honkanen, A. Morsch, G. Rodrigo and J. Schukraft for helpful 
discussions.

%
\appendix

\section{Gluon energy distribution to first order in opacity}
\label{appa}
In this appendix, we give details for the calculation of the zeroth
and first order in opacity of the gluon energy distribution (\ref{2.1}).
We start by expanding the path integral in Eq. (\ref{2.1})  
in powers of $n(\xi)\, \sigma({\bf r})$, 
\begin{eqnarray}
 &&{\cal K}({\bf r},y_l;{\bf \bar r},\bar{y}_l) =
    {\cal K}_0({\bf r},y_l;{\bf \bar r},\bar{y}_l)
 \nonumber \\
 && - \int\limits_{z}^{z'}\, {\it d}\xi\, n(\xi)\, 
 \int {\it d}\bbox{\rho}\,
 {\cal K}_0({\bf r},y_l;\bbox{\rho},\xi)\, \frac{1}{2}\,
   \sigma(\bbox{\rho})\, 
   {\cal K}_0(\bbox{\rho},\xi;{\bf \bar r},\bar{y}_l)
 \nonumber \\
 && + \int\limits_{z}^{z'} {\it d}\xi_1\, n(\xi_1)\, 
    \int\limits_{\xi_1}^{z'} {\it d}\xi_2\, n(\xi_2)\, 
    \int {\it d}\bbox{\rho}_1\,{\it d}\bbox{\rho}_2\,
    {\cal K}_0({\bf r},y_l;\bbox{\rho}_1,\xi_1)\,  
    \nonumber \\
 && \times \frac{1}{2}\,
    \sigma(\bbox{\rho}_1)\, 
    {\cal K}(\bbox{\rho}_1,\xi_1;\bbox{\rho}_2,\xi_2)\, \frac{1}{2}\,
    \sigma(\bbox{\rho}_2)\,
    {\cal K}_0(\bbox{\rho}_2,\xi_2;{\bf \bar r},\bar{y}_l)\, .
 \label{a.1} 
\end{eqnarray}
Here, the free path integral ${\cal K}_0$ is of  
Gaussian form
\begin{equation}
  {\cal K}_0({\bf r},y_l;{\bf \bar r},\bar{y}_l)
  = \frac{\omega}{2\pi\, i\, (\bar{y}_l-y_l)}
    \exp\left\{ { {i\omega}
           \left(\bar{\bf r} - {\bf r}\right)^2 \over {2\, (\bar{y}_l-y_l)} }
           \right\}\, .
  \label{a.2}
\end{equation}
We expand the integrand of (\ref{2.1}) to first order in
$n(\xi)\, \sigma(r)$. The longitudinal integrals in 
(\ref{2.1}) are regularized in intermediate steps of the 
calculation, as explained in Ref.~\cite{Wiedemann:1999fq,Wiedemann:2000za}.
We note that the $N$-th order of (\ref{2.1}) involves $2N + 1$ terms only.
This makes it straightforward to obtain explicit expressions to high 
orders in opacity. 

Inserting (\ref{a.1}) into (\ref{2.1}), one finds to zeroth order
opacity term (\ref{3.1}). To first order in opacity, one
finds 
\begin{eqnarray}
  \omega\frac{dI(N=1)}{d\omega\, d{\bf k}_\perp}
  &=& \frac{\alpha_s\, C_F}{\pi^2} 4\, n_0\, \omega 
    \int \frac{d{\bf q}}{(2\pi)^2}  \vert a({\bf q})\vert^2\, 
    \frac{ LQ_1 - \sin(LQ_1)}{
    \left[ ({\bf k}_\perp + {\bf q})^2 + x^2 m^2 \right]^2}
    \nonumber \\
   && \qquad \times \left[  
     \frac{({\bf k}_\perp + {\bf q})^2}
     {({\bf k}_\perp + {\bf q})^2 + x^2 m^2}
      - \frac{{\bf k}_\perp\cdot ({\bf k}_\perp + {\bf q})}
       { {\bf k}_\perp^2 + x^2 m^2}
   \right]\, ,
   \label{a.3}
\end{eqnarray}
where the transverse energy of the scattered gluon is 
$Q_1 = \frac{({\bf k}_\perp + {\bf q})^2 + x^2 m^2}{2\omega}$.

\underline{Incoherent limit}\\
In the totally incoherent limit, the first order opacity term
(\ref{a.3}) has a probabilistic parton cascade interpretation:
\begin{eqnarray}
  \lim_{L\to\infty}
  \omega\frac{dI(N=1)}{d\omega\, d{\bf k}_\perp} 
  \Bigg\vert_{n_0L = const} 
  &=& \frac{\alpha_s\, C_F}{\pi^2} (n_0\, L)\, 
    \int \frac{d{\bf q}}{(2\pi)^2}  \vert a({\bf q})\vert^2\, 
  \nonumber \\
  && \times \left[ - H({\bf k}_\perp) + H({\bf k}_\perp+{\bf q}) 
                     + R({\bf k}_\perp,{\bf q})  
  \right]\, .
  \label{a.4}
\end{eqnarray}
Here, $H({\bf k}_\perp)$ is multiplied by the probability that the
hard parton interacts with the medium; the minus sign ensures
probability conservation by reducing the corresponding weight of 
the $N=0$ vacuum contribution (\ref{3.1}). For a scattering center 
far away from the point
of quark production, interaction with the medium gives
rise to two processes: i) The vacuum radiation term $H({\bf k}_\perp)$ is
shifted to $H({\bf k}_\perp+{\bf q})$ 
due to medium-induced elastic scattering by a transverse 
momentum ${\bf q}$. ii) Gluons are produced due to bremsstrahlung off
the far distant scattering center. This leads to the Gunion-Bertsch
radiation term for massive quarks,
\begin{eqnarray}
  R({\bf k}_\perp,{\bf q}) &=&   
  \frac{({\bf k}_\perp + {\bf q})^2}{[({\bf k}_\perp + {\bf q})^2 + x^2 m^2]^2}
  - \frac{2\, {\bf k}_\perp\cdot({\bf k}_\perp + {\bf q})}
     {[({\bf k}_\perp + {\bf q})^2 + x^2 m^2]\, [{\bf k}_\perp^2 + x^2 m^2]}
  \nonumber \\
  && + \frac{{\bf k}_\perp^2}{[{\bf k}_\perp^2 + x^2 m^2]^2}\, .
  \label{a.5}
\end{eqnarray}
For realistic, finite in-medium pathlength, both effects combine
in the specific interference pattern (\ref{a.3}).

\underline{$N=1$ opacity term for Yukawa-type scattering potential:}\\
We have studied (\ref{a.3}) for arbitrary in-medium pathlength for a 
Yukawa-type elastic scattering center with Debye screening mass $\mu$:
\begin{equation}
  \vert a({\bf q})\vert^2 = 
  (2\pi)^2\, \frac{\mu^2}{\pi ({\bf q}^2 + \mu^2)^2}\, .
  \label{a.6}
\end{equation}
Shifting in (\ref{a.3}) the relative momentum integration
${\bf q} \to {\bf q} + {\bf k}$ and doing the
angular integrations, one finds
\begin{eqnarray}
  \omega\frac{dI(N=1)}{d\omega\, d{k}^2}
  &=& \frac{\alpha_s\, C_F}{\pi} \frac{n_0\, \mu^2}{\omega}  
    \int_0^\infty d{q}^2   
    \frac{LQ^\prime _1  - \sin(LQ^\prime _1)}{{Q^\prime}_1^2}
    \frac{q^2}{q^2 + x^2 m^2}
    \nonumber \\
   && \qquad \times 
   \frac{\mu^2\left( k^2 + x^2 m^2\right) + 
         \left( k^2 - x^2 m^2 \right) \left( k^2 - q^2 \right)}
       {\left( k^2 + x^2 m^2 \right) 
         \left[ \left(m^2 + k^2 + q^2 \right)^2 - 4 k^2 q^2 \right]^{3/2}}\, ,
   \label{a.7}
\end{eqnarray}
where $Q_1^\prime = \frac{{\bf q}^2 + x^2 m^2}{2 \omega}$.
From this expression,
we find Eq. (\ref{3.14}) by rescaling the variables $\omega$,
$k^2$ and $m^2$ to the dimensionless variables $\bar{\gamma}$,
$\bar{\kappa}^2$ and $\bar{M}^2$ defined in eqs. (\ref{3.4}),
(\ref{3.7}) and (\ref{3.11}) respectively.

\section{Gluon energy distribution in the dipole approximation}
\label{appb}
Here, we give the full expression for the medium-induced part
of the double differential gluon distribution (\ref{4.4}). 
The calculation is done in complete analogy to the calculation
of the massless case~\cite{Wiedemann:2000tf}, but keeping the
mass-dependence of (\ref{2.1}). In the rescaled dimensionless
variables introduced in Section~\ref{sec4}, one finds
\begin{eqnarray}
  \omega \frac{dI_4}{d\omega\, d\kappa^2} &=&
  \frac{\alpha_s\, C_F}{\pi} \gamma^2\, 2\, {\rm Re}
  \int_0^1 dt \int_t^1 d\bar{t}\, e^{i\, M^2\, \gamma\, (t-\bar{t})}
  \exp\left[- \frac{\kappa^2}{4(D_4-iA_4B_4)}\right]
  \nonumber \\
  && \qquad\qquad \times
  \left[ \frac{iA_4^3\, B_4\, \kappa^2}{(D_4-iA_4B_4)^3}
         - \frac{ 4 A_4^2 D_4}{(D_4-i A_4 B_4)^2} \right]\, ,
  \label{b.1}
\end{eqnarray} 
where
\begin{equation}
    \Omega = \frac{1-i}{\sqrt{2}} \sqrt{\gamma}
    \label{b.2}
\end{equation}
and
\begin{eqnarray}
  A_4 = \frac{\Omega}{4 \gamma\, \sin\left[\Omega(\bar{t}-t)\right]}\, ,\quad
  B_4 = \cos\left[\Omega(\bar{t}-t)\right]\, ,\quad
  D_4 = \frac{1}{4} \left( 1 - \bar{t}\right)\, .
  \label{b.3}
\end{eqnarray}
The term $I_5$ in Eq. (\ref{4.2}) takes the form
\begin{eqnarray}
  \omega \frac{dI_5}{d\omega\, d\kappa^2} &=&
  \frac{\alpha_s\, C_F}{\pi} \gamma\, 2\, {\rm Re}
  \int_0^1 dt\, e^{- i\, M^2\, \gamma\, t}
  \left( \frac{-i \kappa^2}{\kappa^2 + M^2}\right)
  \nonumber \\
  && \qquad \qquad \qquad \qquad \times
   \frac{1}{B_5^2}
  \exp\left[- \frac{i \kappa^2}{4\, A_5\, B_5}\right]\, ,
  \label{b.4}
\end{eqnarray} 
where
\begin{eqnarray}
  A_5 = \frac{\Omega}{4 \gamma\, \sin\left[\Omega\, t\right]}\, ,\quad
  B_5 = \cos\left[\Omega\, t\right]\,  .
  \label{b.5}
\end{eqnarray}
%



\begin{references}
%
\bibitem{Dokshitzer:fc}
Y.~L.~Dokshitzer, V.~A.~Khoze and S.~I.~Troian,
J.\ Phys.\ G {\bf 17} (1991) 1481.
%
\bibitem{Ackerstaff:1998hz}
K.~Ackerstaff {\it et al.}  [OPAL Collaboration],
Eur.\ Phys.\ J.\ C {\bf 7} (1999) 369
[arXiv:hep-ex/9807004].
%
\bibitem{Buskulic:1993iu}
D.~Buskulic {\it et al.}  [ALEPH Collaboration],
Z.\ Phys.\ C {\bf 62} (1994) 1.
%
\bibitem{Abbiendi:2002vt}
G.~Abbiendi {\it et al.}  [OPAL Collaboration],
arXiv:hep-ex/0210031.

\bibitem{Schumm:1992xt}
B.~A.~Schumm, Y.~L.~Dokshitzer, V.~A.~Khoze and D.~S.~Koetke,
Phys.\ Rev.\ Lett.\  {\bf 69} (1992) 3025.
%
\bibitem{Gieseke:2003rz}
S.~Gieseke, P.~Stephens and B.~Webber,
arXiv:hep-ph/0310083.
%
\bibitem{Gyulassy:1993hr}
M.~Gyulassy and X.~N.~Wang,
Nucl.\ Phys.\ B {\bf 420} (1994) 583
[arXiv:nucl-th/9306003].
%
\bibitem{Baier:1996sk}
R.~Baier, Y.~L.~Dokshitzer, A.~H.~Mueller, S.~Peigne and D.~Schiff,
Nucl.\ Phys.\ B {\bf 484} (1997) 265
[arXiv:hep-ph/9608322].
%
\bibitem{Zakharov:1997uu}
B.~G.~Zakharov,
JETP Lett.\  {\bf 65} (1997) 615
[arXiv:hep-ph/9704255].
%
\bibitem{Wiedemann:2000za}
U.~A.~Wiedemann,
Nucl.\ Phys.\ B {\bf 588} (2000) 303
[arXiv:hep-ph/0005129].
%
\bibitem{Gyulassy:2000er}
M.~Gyulassy, P.~Levai and I.~Vitev,
Nucl.\ Phys.\ B {\bf 594} (2001) 371
[arXiv:nucl-th/0006010].
%
\bibitem{Wang:2001if}
X.~N.~Wang and X.~f.~Guo,
Nucl.\ Phys.\ A {\bf 696} (2001) 788
[arXiv:hep-ph/0102230].
%
\bibitem{Adcox:2001jp}
K.~Adcox {\it et al.}  [PHENIX Collaboration],
Phys.\ Rev.\ Lett.\  {\bf 88} (2002) 022301
[arXiv:nucl-ex/0109003].
%
\bibitem{Adler:2003au}
S.~S.~Adler  [PHENIX Collaboration],
arXiv:nucl-ex/0308006.
%
\bibitem{Adler:2002xw}
C.~Adler {\it et al.} [STAR Collaboration],
Phys.\ Rev.\ Lett.\  {\bf 89} (2002) 202301
[arXiv:nucl-ex/0206011].
%
\bibitem{Adams:2003kv}
J.~Adams {\it et al.}  [STAR Collaboration],
Phys.\ Rev.\ Lett.\  {\bf 91} (2003) 172302
[arXiv:nucl-ex/0305015].
%
\bibitem{Back:2003qr}
B.~B.~Back {\it et al.}  [PHOBOS Collaboration],
arXiv:nucl-ex/0302015.
%
\bibitem{Arsene:2003yk}
I.~Arsene {\it et al.}  [BRAHMS Collaboration],
Phys.\ Rev.\ Lett.\  {\bf 91} (2003) 072305
[arXiv:nucl-ex/0307003].
%
\bibitem{Wang:2003aw}
X.~N.~Wang,
arXiv:nucl-th/0307036.
%
\bibitem{Wiedemann:2000tf}
U.~A.~Wiedemann,
Nucl.\ Phys.\ A {\bf 690} (2001) 731
[arXiv:hep-ph/0008241].
%
\bibitem{Baier:1999ds}
R.~Baier, Y.~L.~Dokshitzer, A.~H.~Mueller and D.~Schiff,
Phys.\ Rev.\ C {\bf 60} (1999) 064902
[arXiv:hep-ph/9907267].
%
\bibitem{Baier:2001qw}
R.~Baier, Y.~L.~Dokshitzer, A.~H.~Mueller and D.~Schiff,
Phys.\ Rev.\ C {\bf 64} (2001) 057902
[arXiv:hep-ph/0105062].
%
\bibitem{Salgado:2003gb}
C.~A.~Salgado and U.~A.~Wiedemann,
Phys.\ Rev.\ D {\bf 68} (2003) 014008
[arXiv:hep-ph/0302184].
%
\bibitem{Salgado:2003rv}
C.~A.~Salgado and U.~A.~Wiedemann,
arXiv:hep-ph/0310079.
%
\bibitem{Dokshitzer:2001zm}
Yu.~L.~Dokshitzer and D.~E.~Kharzeev,
Phys.\ Lett.\ B {\bf 519} (2001) 199
[arXiv:hep-ph/0106202].
%
\bibitem{Eskola:2003fk}
K.~J.~Eskola, V.~J.~Kolhinen and R.~Vogt,
arXiv:hep-ph/0310111.
%
\bibitem{Gelis:2003vh}
F.~Gelis and R.~Venugopalan,
arXiv:hep-ph/0310090.
%
\bibitem{Kharzeev:2003sk}
D.~Kharzeev and K.~Tuchin,
arXiv:hep-ph/0310358.
%
\bibitem{Accardi:2003be}
A.~Accardi {\it et al.},
arXiv:hep-ph/0308248.
%
\bibitem{Djordjevic:2003qk}
M.~Djordjevic and M.~Gyulassy,
Phys.\ Lett.\ B {\bf 560} (2003) 37
[arXiv:nucl-th/0302069].
%
\bibitem{Djordjevic:2003be}
M.~Djordjevic and M.~Gyulassy,
Phys.\ Rev.\ C {\bf 68} (2003) 034914
[arXiv:nucl-th/0305062].
%
\bibitem{Djordjevic:2003zk}
M.~Djordjevic and M.~Gyulassy,
arXiv:nucl-th/0310076.
%
\bibitem{Zhang:2003wk}
B.~W.~Zhang, E.~Wang and X.~N.~Wang,
arXiv:nucl-th/0309040.
%
\bibitem{Zakharov:1996fv}
B.~G.~Zakharov,
JETP Lett.\  {\bf 63} (1996) 952
[arXiv:hep-ph/9607440].
%
\bibitem{Wiedemann:1999fq}
U.~A.~Wiedemann and M.~Gyulassy,
Nucl.\ Phys.\ B {\bf 560} (1999) 345
[arXiv:hep-ph/9906257].
%
\bibitem{Salgado:2002cd}
C.~A.~Salgado and U.~A.~Wiedemann,
Phys.\ Rev.\ Lett.\  {\bf 89} (2002) 092303
[arXiv:hep-ph/0204221].
%
\bibitem{Gyulassy:2000fs}
M.~Gyulassy, P.~Levai and I.~Vitev,
Phys.\ Rev.\ Lett.\  {\bf 85} (2000) 5535
[arXiv:nucl-th/0005032].
%
\bibitem{Levai:2001dc}
P.~Levai, G.~Papp, G.~Fai, M.~Gyulassy, G.~G.~Barnafoldi, I.~Vitev 
and Y.~Zhang,
Nucl.\ Phys.\ A {\bf 698} (2002) 631
[arXiv:nucl-th/0104035].
%
\bibitem{Zakharov:1998sv}
B.~G.~Zakharov,
Phys.\ Atom.\ Nucl.\  {\bf 61} (1998) 838
[Yad.\ Fiz.\  {\bf 61} (1998) 924]
[arXiv:hep-ph/9807540].
%
\bibitem{Baier:2002tc}
R.~Baier,
Nucl.\ Phys.\ A {\bf 715} (2003) 209
[arXiv:hep-ph/0209038].
%
\bibitem{Shuryak:1996gc}
E.~V.~Shuryak,
Phys.\ Rev.\ C {\bf 55} (1997) 961
[arXiv:nucl-th/9605011].
%
\bibitem{Lin:1997cn}
Z.~w.~Lin, R.~Vogt and X.~N.~Wang,
Phys.\ Rev.\ C {\bf 57} (1998) 899
[arXiv:nucl-th/9705006].
%
\bibitem{Mustafa:1997pm}
M.~G.~Mustafa, D.~Pal, D.~K.~Srivastava and M.~Thoma,
Phys.\ Lett.\ B {\bf 428} (1998) 234
[arXiv:nucl-th/9711059].
%
\bibitem{Adcox:2002cg}
K.~Adcox {\it et al.}  [PHENIX Collaboration],
Phys.\ Rev.\ Lett.\  {\bf 88} (2002) 192303
[arXiv:nucl-ex/0202002].
%
\bibitem{Dainese:2003wq}
A.~Dainese,
arXiv:nucl-ex/0312005.
%
\bibitem{Lokhtin:ay}
I.~P.~Lokhtin and A.~M.~Snigirev,
J.\ Phys.\ G {\bf 27} (2001) 2365.

\end{references}
\end{document}